# Fluctuation Electromagnetic Interaction Between Small Rotating Particle and a Surface

G.V. Dedkov and A.A. Kyasov

Nanoscale Physics Group, Kabardino-Balkarian State University, 360004, Nalchik, Russia

We study fluctuation electromagnetic interaction between small neutral rotating particle and polarizable surface. The attraction force, friction torque and heating are produced by the particle polarization and fluctuating near-field of the surface. Closed analytical expressions are found for these quantities assuming that the particle and surface are characterized by the frequency dependent polarizability and dielectric permittivity and different temperatures. It is shown that the stopping time of the rolling particle in the near-field of the surface is much less than the stopping time under uniform motion and rolling in vacuum space.

## 1. Introduction

The near-field fluctuation electromagnetic interaction between a particle in uniform motion and the surface is the subject of continuing interest demonstrating a lot of new features relative to the simplest case of resting particle. In particular, the origin of noncontact friction between the bodies in relative motion has been debated for a long time [1] (see also recent works [2-5]). Quite recently Manjavacas and Abajo [6] considered the problem of friction torque acting on a particle rotating in vacuum. The friction torque is produced by fluctuations of the vacuum electromagnetic field and the particle polarization. A related instance is the friction and heating of the particle moving at a constant velocity in the vacuum background [7, 8].

Rotation of the particle leads to the frequency shifts in fluctuating fields involving the rotation frequency $\Omega$. The attraction, friction and heating between rotating bodies will therefore be different in comparison with the static case. The aim of this paper is to examine the general problem of fluctuation electromagnetic interaction between the spherical rotating particle and the surface, using the standard formalism of fluctuating electrodynamics [9].

## 2. Theory

The system under consideration is shown in Fig. 1, which consists of an isotropic particle at temperature $T_1$ placed in vacuum at a distance $z_0$ from the surface at temperature $T_2$. The particle rotates around its $z$ axis with the angular frequency $\Omega$ assuming to be a point-like fluctuating nonrelativistic dipole. In this case the following conditions are fulfilled ($\omega_0$ is the characteristic absorption frequency of the particle, $c, k_B,$ and $\hbar$ are the speed of light in vacuum, the Boltzmann and Planck's constants)

$$R \ll \min\left\{\frac{2\pi c}{\omega_0}, \frac{2\pi c}{\Omega}, \frac{2\pi \hbar c}{k_B T_1}, \frac{2\pi \hbar c}{k_B T_2}\right\}$$

Apart from the force of attraction to the surface $F_z$ and the rate of heat exchange $\dot{Q}$, the interaction produces frictional torque $M$ on the particle. All these quantities are calculated within a unified approach based on fluctuation electrodynamics that we have first proposed in [10] ( for a review see [11]; general relativistic results see in [2]). According to our method, we consider the contribution of all independent sources of electromagnetic fluctuations, namely $\mathbf{d}^{sp}$ and $\mathbf{E}^{sp}$ in this case, where the former and latter quantities are the fluctuating dipole moment of the particle and fluctuating electromagnetic field of the surface.

## 2.1 Frictional torque

The torque is given by

$$M_z = \langle [\mathbf{dE}]_z \rangle = \langle [\mathbf{d}^{sp}\mathbf{E}^{in}]_z \rangle + \langle [\mathbf{d}^{in}\mathbf{E}^{in}]_z \rangle = M_z^{(1)} + M_z^{(2)} \qquad (1)$$

where $\langle ... \rangle$ denotes total quantum-statistical averaging, the superscripts "*sp*" and "*in*" denote spontaneous and induced quantities taken in the resting coordinate system $\Sigma$ $(x, y, z)$ related to the surface. The contribution $M_z^{(1)}$ is given by

$$M_z^{(1)} = \langle d_x^{sp} E_y^{in} - d_y^{sp} E_x^{in} \rangle \qquad (2)$$

Using the Fourier transforms

$$\mathbf{d}^{sp}(t) = \int_{-\infty}^{+\infty} \frac{d\omega}{2\pi} \mathbf{d}^{sp}(\omega) \exp(-i\omega t) \qquad (3)$$

$$\mathbf{E}^{in}(t) = \int \frac{d\omega}{2\pi} \frac{d^2 k}{(2\pi)^2} \mathbf{E}^{in}(\omega \mathbf{k}, z) \exp(i(k_x x + k_y y - \omega t)) \qquad (4)$$

we take Eq.(4) at the particle location point $\mathbf{r}_0 = (0,0,z_0)$. Inserting (3),(4) into (2) we then obtain

$$M_z^{(1)} = \int \frac{d\omega'}{2\pi} \frac{d\omega}{2\pi} \frac{d^2k}{(2\pi)^2} \exp(-i(\omega+\omega')t) \langle d_x^{sp}(\omega') E_y^{in}(\omega\mathbf{k}, z_0) - d_y^{sp}(\omega') E_x^{in}(\omega\mathbf{k}, z_0) \rangle \qquad (5)$$

Using equation $\mathbf{E}^{in} = -\nabla \phi^{in}$, the components $\mathbf{E}^{in}(\omega\mathbf{k}, z)$ are expressed through the Fourier components of the induced potential of the surface [11]

$$\phi^{in}(\omega\mathbf{k}, z) = \frac{2\pi}{k} \Delta(\omega) \exp(-k(z+z_0)) \left[ ik_x d_x^{sp}(\omega) + ik_y d_y^{sp}(\omega) + k d_z^{sp}(\omega) \right] \qquad (6)$$

$$E_x^{in}(\omega\mathbf{k}, z) = -ik_x \phi^{in}(\omega\mathbf{k}, z),\ E_y^{in}(\omega\mathbf{k}, z) = -ik_y \phi^{in}(\omega\mathbf{k}, z),\ E_z^{in}(\omega\mathbf{k}, z) = k\phi^{in}(\omega\mathbf{k}, z) \qquad (7)$$

$$\Delta(\omega) = \frac{\varepsilon(\omega) - 1}{\varepsilon(\omega) + 1} \qquad (8)$$

where $\varepsilon(\omega)$ is the dielectric permittivity of the surface and $k = |\mathbf{k}| = (k_x^2 + k_y^2)^{1/2}$.

The fluctuation-dissipation theorem (FDT) for the dipole operator $\mathbf{d}^{sp}$ is given in the rotating coordinate system $\Sigma'$ of the particle

$$\langle d_i^{sp'}(\omega') d_k^{sp'}(\omega) \rangle = 2\pi \delta_{ik} \delta(\omega + \omega') \hbar \alpha''(\omega) \coth \frac{\hbar\omega}{2k_B T_1} \qquad (9)$$

where $T_1$ and $\alpha(\omega)$ are the particle temperature and polarizability in the coordinate system $\Sigma'$ ($\alpha''(\omega)$ is the corresponding imaginary component). In order to calculate the correlator in Eq. (5) using FDT (9), we must express the Fourier transforms of $\mathbf{d}^{sp}$ in $\Sigma$ through the Fourier transforms of $\mathbf{d}^{sp'}$ in $\Sigma'$. Any vector quantity $\mathbf{A}$ in $\Sigma$ is related to $\mathbf{A}'$ in $\Sigma'$ by

$$\begin{pmatrix} A_x \\ A_y \\ A_z \end{pmatrix} = \begin{pmatrix} \cos\Omega t & -\sin\Omega t & 0 \\ \sin\Omega t & \cos\Omega t & 0 \\ 0 & 0 & 1 \end{pmatrix} \begin{pmatrix} A_x' \\ A_y' \\ A_z' \end{pmatrix} \qquad (10)$$

Using (10), the Fourier components of $\mathbf{d}^{sp}$ in $Z$ are given by

$$d_x^{sp}(t) = \cos\Omega t \cdot d_x^{sp'}(t) - \sin\Omega t \cdot d_y^{sp'}(t)$$

$$d_y^{sp}(t) = \sin\Omega t \cdot d_x^{sp'}(t) + \cos\Omega t \cdot d_y^{sp'}(t)$$

$$d_z^{sp}(t) = d_x^{sp'}(t) \qquad (11)$$

where $\Omega$ is the angular rotation frequency of the system $\Sigma'$ with respect to $\Sigma$. The Fourier transforms of (11) are given by

$$d_x^{sp}(\omega) = \frac{1}{2} \left( d_x^{sp'}(\omega^+) + d_x^{sp'}(\omega^-) + id_y^{sp'}(\omega^+) - id_y^{sp'}(\omega^-) \right)$$

$$d_y^{sp}(\omega) = \frac{1}{2}\left(-id^{sp}{}_x{}'(\omega^+) + id^{sp}{}_x{}'(\omega^-) + d^{sp}{}_y{}'(\omega^+) + d^{sp}{}_y{}'(\omega^-)\right)$$

$$d_z^{sp}(\omega) = d_z^{sp}{}'(\omega) \tag{12}$$

where $\omega^\pm = \omega \pm \Omega$.

Using FDT (9) and inserting Eqs. (6),(7),(12) in (5) yields

$$M_z^{(1)} = -\frac{\hbar}{4\pi z_0^3}\int_0^\infty d\omega \Delta''(\omega)\left[\alpha''(\omega^-)\coth\frac{\hbar\omega^-}{2k_B T_1} - \alpha''(\omega^+)\coth\frac{\hbar\omega^+}{2k_B T_1}\right] \tag{13}$$

Now we pass to the calculation of $M_z^{(2)}$ in Eq. (1) that is given by

$$M_z^{(2)} = \langle d_x^{in} E_y^{sp} - d_y^{in} E_x^{sp} \rangle \tag{14}$$

Substituting the frequency Fourier expansions of $\mathbf{d}^{in}(t)$ and $\mathbf{E}^{sp}$ in (14) yields

$$M_z^{(2)} = \int\frac{d\omega'}{2\pi}\frac{d\omega}{2\pi}\exp(-i(\omega+\omega')t)\langle d_x^{in}(\omega)E_y^{sp}(\mathbf{r}_0,\omega') - d_y^{in}(\omega)E_x^{sp}(\mathbf{r}_0,\omega')\rangle \tag{15}$$

In the rotating system $\Sigma'$ of the particle we obviously have the relationships

$$d^{in}{}_x{}'(\omega) = \alpha(\omega)E^{sp}{}_x{}'(\mathbf{r}_0,\omega)$$
$$d^{in}{}_y{}'(\omega) = \alpha(\omega)E^{sp}{}_y{}'(\mathbf{r}_0,\omega), \tag{16}$$
$$d^{in}{}_z{}'(\omega) = \alpha(\omega)E^{sp}{}_z{}'(\mathbf{r}_0,\omega)$$

Since the induced components of $\mathbf{d}^{in}$ are given by the equations quite analogous to (12), we obtain with allowance for (16) (for brevity, we omit the argument $\mathbf{r}_0$ in what follows)

$$d_x^{in}(\omega) = \frac{1}{2}\left(\alpha(\omega^+)E^{sp}{}_x{}'(\omega^+) + \alpha(\omega^-)E^{sp}{}_x{}'(\omega^-) + i\alpha(\omega^+)E^{sp}{}_y{}'(\omega^+) - i\alpha(\omega^-)E^{sp}{}_y{}'(\omega^-)\right)$$

$$d_y^{in}(\omega) = \frac{1}{2}\left(-i\alpha(\omega^+)E^{sp}{}_x{}'(\omega^+) + i\alpha(\omega^-)E^{sp}{}_x{}'(\omega^-) + \alpha(\omega^+)E^{sp}{}_y{}'(\omega^+) + \alpha(\omega^-)E^{sp}{}_y{}'(\omega^-)\right)$$

$$d_z^{in}(\omega) = \alpha(\omega)E_z^{sp}{}'(\omega) \tag{17}$$

In addition, using Eq. (10) we express the components of $\mathbf{E}^{sp}{}'$ in $\Sigma'$ through the components of $\mathbf{E}^{sp}$ in $\Sigma$. The resulting formulas are given by

$$E^{sp}_x{}'(\omega) = \frac{1}{2}\left[E^{sp}_x(\omega^+) + E^{sp}_x(\omega^-) - iE^{sp}_y(\omega^+) + iE^{sp}_y(\omega^-)\right]$$

$$E^{sp}_y{}'(\omega) = \frac{1}{2}\left[iE^{sp}_x(\omega^+) - iE^{sp}_x(\omega^-) + E^{sp}_y(\omega^+) + E^{sp}_y(\omega^-)\right] \qquad (18)$$

$$E^{sp}_z{}'(\omega) = E^{sp}_z(\omega)$$

Substituting (18) in (17) yields

$$d^{in}_x(\omega) = \frac{1}{2}\left[\alpha(\omega^+)\left(E^{sp}_x(\omega) + iE^{sp}_y(\omega)\right) + \alpha(\omega^-)\left(E^{sp}_x(\omega) - iE^{sp}_y(\omega)\right)\right]$$

$$d^{in}_y(\omega) = \frac{1}{2}\left[-i\alpha(\omega^+)\left(E^{sp}_x(\omega) + iE^{sp}_y(\omega)\right) + i\alpha(\omega^-)\left(E^{sp}_x(\omega) - iE^{sp}_y(\omega)\right)\right]$$

$$d^{in}_z(\omega) = \alpha(\omega)E^{sp}_z(\omega) \qquad (19)$$

Using (15) with allowance for (19) we obtain

$$M_z^{(2)} = \frac{i}{2}\int \frac{d\omega'}{2\pi}\frac{d\omega}{2\pi}\exp(-i(\omega+\omega')t)(\alpha(\omega^+) - \alpha(\omega^-))\cdot$$
$$\cdot\left[\langle E^{sp}_x(\omega)E^{sp}_x(\omega')\rangle + \langle E^{sp}_y(\omega)E^{sp}_y(\omega')\rangle\right]. \qquad (20)$$

Correlators for the electric–field fluctuations in Eq. (20) are expressed through the imaginary part of retarded Green's function and in configuration under consideration are given by [11]

$$\langle E^{sp}_i(\omega)E^{sp}_i(\omega')\rangle = 2\pi\delta(\omega+\omega')\left(E^{sp}_i\right)^2_\omega, \quad i = x, y, z \qquad (21)$$

$$\left(E^{sp}_i\right)^2_\omega = \int \frac{d^2k}{(2\pi)^2}\left(E^{sp}_i\right)^2_{\omega\mathbf{k}} \qquad (22)$$

$$\left(E^{sp}_x\right)^2_{\omega\mathbf{k}} = \hbar\coth\frac{\hbar\omega}{2k_BT_2}\Delta''(\omega)\frac{2\pi}{k}k_x^2\exp(-2kz_0) \qquad (23)$$

$$\left(E^{sp}_y\right)^2_{\omega\mathbf{k}} = \hbar\coth\frac{\hbar\omega}{2k_BT_2}\Delta''(\omega)\frac{2\pi}{k}k_y^2\exp(-2kz_0) \qquad (24)$$

$$\left(E^{sp}_z\right)^2_{\omega\mathbf{k}} = \hbar\coth\frac{\hbar\omega}{2k_BT_2}\Delta''(\omega)\frac{2\pi}{k}k^2\exp(-2kz_0) \qquad (25)$$

where $T_2$ is the surface temperature. Equations (21)—(25) define the spectral density of the fluctuating electromagnetic field at the particle location point $\mathbf{r}_0 = (0,0,z_0)$. Substituting Eqs.(21)—(25) in (20) and making use simple transformations yields

$$M_z^{(2)} = -\frac{\hbar}{4\pi z_0^3} \int_0^\infty d\omega \Delta''(\omega) \left[ \alpha''(\omega^+) \coth\frac{\hbar\omega}{2k_B T_2} - \alpha''(\omega^-) \coth\frac{\hbar\omega}{2k_B T_2} \right] \qquad (26)$$

Finally, the torque resulting from the sum of (13) and (26) reduces to

$$M_z = -\frac{\hbar}{4\pi z_0^3} \int_0^\infty d\omega \Delta''(\omega) \cdot \left\{ \begin{array}{l} \alpha''(\omega^-)\left[\coth\dfrac{\hbar\omega^-}{2k_B T_1} - \coth\dfrac{\hbar\omega}{2k_B T_2}\right] - \\ -\alpha''(\omega^+)\left[\coth\dfrac{\hbar\omega^+}{2k_B T_1} - \coth\dfrac{\hbar\omega}{2k_B T_2}\right] \end{array} \right\} \qquad (27)$$

Transforming the integral limits in Eq. (27) we obtain a more compact expression

$$M_z = \frac{\hbar}{4\pi z_0^3} \int_{-\infty}^{+\infty} d\omega \Delta''(\omega)\alpha''(\omega^+)\left[\coth\frac{\hbar\omega^+}{2k_B T_1} - \coth\frac{\hbar\omega}{2k_B T_2}\right] \qquad (28)$$

In the limit $\Omega \ll \omega_0$ and $T_1 = T_2 = T$, the dominating term of Eq. (28) which is linear in $\Omega$ takes the form

$$M_z = -\frac{\hbar\Omega}{2\pi z_0^3} \int_{-\infty}^{+\infty} d\omega \Delta''(\omega)\alpha''(\omega)\left(-\frac{\partial}{\partial\omega}\right)\coth\frac{\hbar\omega}{2k_B T} \qquad (30)$$

*2.2 Attraction force and the rate of heat exchange*

The particle-surface attraction force and the rate of heat exchange can be calculated quite analogously to the calculation of $M_z$. The starting expressions are given by

$$F_z = \langle \partial_z (\mathbf{d}^{sp}\mathbf{E}^{in}) \rangle + \langle \partial_z (\mathbf{d}^{in}\mathbf{E}^{sp}) \rangle \qquad (31)$$

$$\dot{Q} = \langle \dot{\mathbf{d}}^{sp}\mathbf{E}^{in} \rangle + \langle \dot{\mathbf{d}}^{in}\mathbf{E}^{sp} \rangle \qquad (32)$$

The resulting equations have the form

$$F_z = -\frac{3\hbar}{16\pi z_0^4} \int_{-\infty}^{+\infty} d\omega \left\{ \begin{array}{l} \Delta'(\omega)\alpha''(\omega)\coth\frac{\hbar\omega}{2k_B T_1} + \Delta''(\omega)\alpha'(\omega)\coth\frac{\hbar\omega}{2k_B T_2} + \\ + \Delta'(\omega)\alpha''(\omega^+)\coth\frac{\hbar\omega^+}{2k_B T_1} + \Delta''(\omega)\alpha'(\omega^+)\coth\frac{\hbar\omega}{2k_B T_2} \end{array} \right\} \quad (33)$$

$$\dot{Q} = \frac{\hbar}{8\pi z_0^3} \int_{-\infty}^{+\infty} d\omega\omega \left\{ \begin{array}{l} \Delta''(\omega)\alpha''(\omega)\left[\coth\frac{\hbar\omega}{2k_B T_2} - \coth\frac{\hbar\omega}{2k_B T_1}\right] + \\ \Delta''(\omega)\alpha''(\omega^+)\left[\coth\frac{\hbar\omega}{2k_B T_2} - \coth\frac{\hbar\omega^+}{2k_B T_1}\right] \end{array} \right\} \quad (34)$$

*2.2 Evaluation of stopping times of nanoparticles*

First, it is interesting to compare the stopping time of small rolling particles with that at uniform motion with the constant velocity $V$ parallel to the surface. Assuming the temperatures of the particle and surface to be $T$, the stopping force is given by [11]

$$F_x = -\frac{3\hbar V}{4\pi z_0^5} \int_{-\infty}^{+\infty} d\omega \Delta''(\omega)\alpha''(\omega)\left(-\frac{\partial}{\partial\omega}\right)\coth\frac{\hbar\omega}{2k_B T} \quad (35)$$

Using (30), (35) and Newton's second law, the corresponding stopping times (assuming that the velocity diminishes by $e$ times) are given by $\tau_\Omega = \frac{4\pi}{5\hbar J}mR^2 z_0^3$ and $\tau_V = \frac{4\pi}{3\hbar J}mz_0^5$, where $m$ and $R$ are the particle mass and radius, and $J$ is the frequency integral in (30) and (35). We also took into account the inertia moment of spherical particle which is equal to $2mR^2/5$. Using the above expressions we obtain $\tau_\Omega/\tau_V = \frac{3}{5}\left(\frac{R}{z_0}\right)^2$. Since typically $R \ll z_0$, this implies that $\tau_\Omega/\tau_V \ll 1$ and the decay time at rolling is much shorter than at a uniform motion.

Second, it is interesting to compare $\tau_\Omega$ near the surface and in the vacuum background (at equal temperature $T$ of the particle and background radiation). In the latter case the friction torque is given by the formula [6] which can be written similar to (30) and (35)

$$M_z^{(vac)} = -\frac{2\hbar\Omega}{3\pi c^3} \int_{-\infty}^{+\infty} d\omega\omega^3 \Delta''(\omega)\alpha''(\omega)\left(-\frac{\partial}{\partial\omega}\right)\coth\frac{\hbar\omega}{2k_B T} \quad (36)$$

Using (36), the stopping time is $\tau_\Omega^{(vac)} = \frac{6\pi}{5\hbar J_{vac}} mR^2 c^3$, where $J_{vac}$ is the frequency integral in Eq. (36). Hence it follows $\tau_\Omega / \tau_\Omega^{(vac)} = \frac{2}{3} \frac{J_{vac}}{J} \left(\frac{z_0}{c}\right)^3$. Comparing the structure of $J_{vac}$ and $J$, we can conclude that $J_{vac}/J \propto (k_B T/\hbar)^3 = \omega_W^3$ (with $\omega_W$ being the Wien frequency). Therefore, ignoring the numerical factor of $1 \div 10$, we obtain $\tau_\Omega / \tau_\Omega^{(vac)} \approx (\omega_W z_0/c)^3$. At room temperature and $z_0 = 100\,nm$, for example, $\tau_\Omega / \tau_\Omega^{(vac)} \approx 2 \cdot 10^{-6}$. So, the stopping time for rolling in the near-field of the surface is much shorter than in vacuum irrespectively of material properties. For metallic particles, the dominating contribution in Eq. (36) results from the magnetic polarizability of the particle [6], that can be higher by $1 \div 2$ orders of magnitude than the contribution of electric polarizability. But generally, the estimation $\tau_\Omega / \tau_\Omega^{(vac)} \ll 1$ proves to be valid in this case, as well.

## 3. Conclusions

Using the general background of fluctuation electromagnetic theory, we have obtained closed nonrelativistic expressions for the friction torque, attraction force and heating rate of a small spherical particle rotating in the near-field of the surface. The material properties of particle and surface are characterized by the frequency-dependent polarizability and dielectric permittivity. The temperatures of the particle and the surface are assumed to be arbitrary. It is worth noting that the anisotropy of the particle polarizability has no appreciable physical importance in this case and may only change the numerical factors in the quantities under study. Apparently, it is not difficult to generalize the obtained formulas in this case.

Assuming isothermal conditions, we have compared the stopping times of spherical particles corresponding to the rolling and uniform motion near a surface and those for rolling near the surface and in the vacuum background. In both of the cases (at room and lower temperatures) the stopping times for rolling in the near-field of the surface turns out to be smaller by several orders of magnitude than in vacuum and at a uniform motion near a surface.

## References

... 
[1] K.A. Milton, Am. J. Phys., 79, 697 (2011).

[2] G.V. Dedkov, A.A. Kyasov, J. Phys.: Condens. Matter 20, 354006 (2008); Surf. Sci. 604, 561 (2010); arXiv: 1112.5619.

[3] T.G. Philbin and U. Leonhardt, New J. Phys. 11, 033035(2009); ibid. 12, 068001(2010).

[4] J.B. Pendry, New J. Phys. 12, 033028 (2010); ibid. 12, 068002(2010).

[5] J.S. Hoye and I. Brevik, Eur. Phys. J. D64, 1(2011); arXiv: 111.4858; arXiv: 1201.3830.

[6] A. Manjavacas and F.J. Garcia de Abajo, Phys. Rev. A82, 063827(2010).

[7] V. Mkrtchian, V.A. Parsegian, R.Podgornik, W.M. Saslov, Phys. Rev. Lett. 91, 220801(2003).

[8] G.V. Dedkov and A.A. Kyasov, Phys. Lett. A339, 212(2005).

[9] S.M. Rytov, Theory of Electromagnetic Fluctuations and Thermal Radiation, Air Force Cambridge research Center, Bedford, MA, 1959, AFCRC-TR-59-162.

[10] G.V. Dedkov and A.A. Kysov, Phys. Lett. A259, 38(1999); Surf. Sci. 463, 11(2000)

[11] G.V. Dedkov and A.A. Kyasov, Phys. Solid State, 44, 1729(2002); Phys. Low.-Dim. Struct. 1/2, 1(2003).

Fig. 1

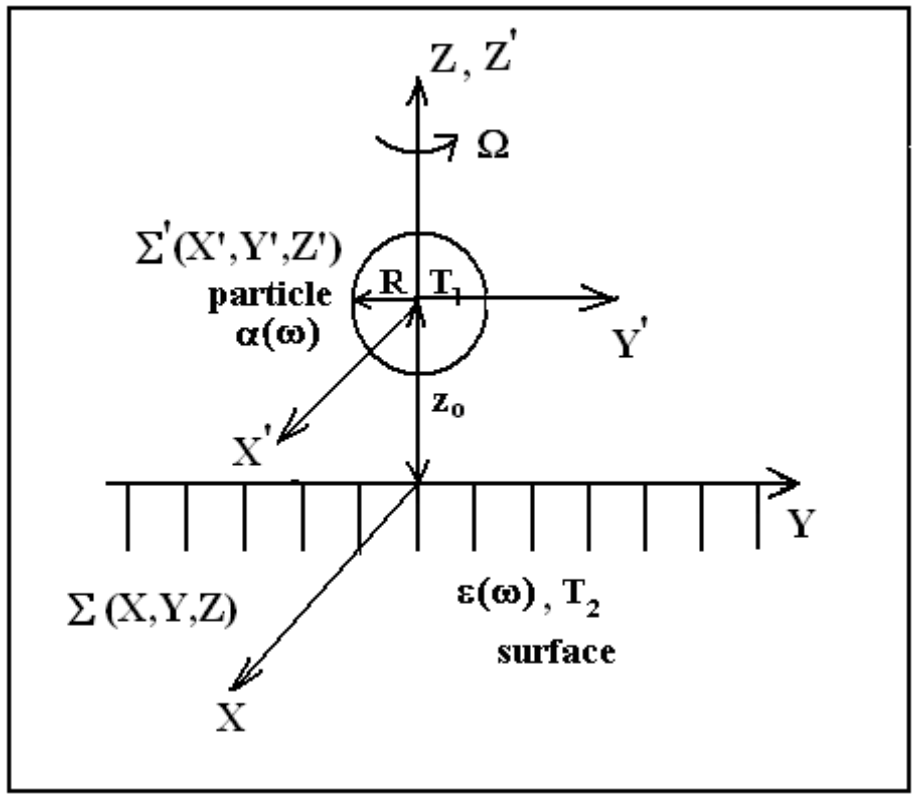

Fig. 1. Coordinate systems used and statement of the problem